# Phasing a deployable sparse telescope

## Erez N. Ribak,[1*] and B. Martin Levine [1]

[1] *Physics Department and Asher Space Research Institute, Technion – Israel Institute of Technology, Haifa 32000, Israel*

*Corresponding author: eribak@physics.technion.ac.il*

**ABSTRACT**

After launching and deploying a sparse space telescope, fine tuning is required to correct for inaccurate initial placement of its elements. We selected unique shapes and locations of these telescope aperture segments, to be able to distinguish between their diffraction patterns, while at the same time having a proper spatial frequency coverage. Then we improved the combined wave front, without measuring it directly: First we correlated each segment's focal image with its distinctive template, to correct its tilt. Next we interfered them with the other segments, pair by pair, using their limited coherence, to locate their mutual optical path differences. Finally, we optimized the combined focal image for fine alignment.

There are strong constraints on the volume and mass of space telescopes, mostly caused by the limited capacity of current launch vehicles. Ground-based telescopes suffer from similar limitations, but to a much lesser degree. The solution to the size limitation in both cases can be segmented telescopes, such as the (original) Multi-Mirror Telescope [1], Keck telescope, and their successors. These segmented telescopes also keep the sectors very close to each other, to avoid background radiation from the gaps, and to maximize surface area. On the other extreme, interferometers gain in resolution by increasing the linear size of the aperture and by making it sparse, even extremely sparse (in the radio regime), all at the price of signal lost between the elements.

If it is not easy to perfectly align any telescope, the task becomes tougher when the telescope is segmented and sparse. Instead of worrying about the limited degrees of freedom (DOFs) of the few elements, now each added segment contributes its own DOFs, anywhere between one and six, excluding optional sub-segment correction (such as curvature adjustment or higher order errors). The accuracy required is that of a contiguous telescope, traditionally a fraction of the shortest wave length.

A large body of methods have already been devised to solve these issues. They are roughly divided between direct and indirect wave front sensing (WFS). WFS usually entails using additional sensors and tools to measure - and correct - the wave front of the segmented telescope or interferometer. Hartmann-Shack, Pyramid, or physical sensors allow bridging measurements of segment-to-segment error, albeit limited to the edge accuracy [2]. Unfortunately, they are less practical for sparse apertures with large gaps.

Indirect WFS is another method, which does not require additional hardware, a significant advantage in space astronomy, where simplicity reduces the chance of equipment failure. It entails measuring the focal intensity distribution, as opposed to direct WFS, performed in the aperture plain. The notion is that less aberrated wave fronts produce images closer to the diffraction limit, namely sharper images, by Parseval's theorem [3, 4]. How each DOF affects the image sharpness is not directly known, so optimisation in hardware seems a straightforward way to achieve wave front control. On smaller systems this optimisation is very easy, but for segmented telescopes with many DOFs, the size of the search volume becomes overly large and unfathomable. Using indirect WFS such as simulated annealing [5] is safe and slow. Unfortunately, when combined with faster steepest descent [6, 7] the search tends to sink into sub-optimal solutions. It was proposed to use a supervised learning algorithm [8], where a full search must first be performed over the same huge exploration volume ahead of time, and memorized for a one-time application. Phase diversity WFS iteratively uses a focal image and an extra-focal one, and works best for compact, narrow-band objects, and contiguous apertures [9]. The solutions that we propose below pertains to segmented space telescopes. In some cases, they can be applied to ground-based telescopes, where turbulence effects might be mitigated by temporal or spatial filtering [10].

Figure 1 shows us the basic idea of a small deployable telescope in the Fresnel configuration [11]. It has two drawbacks: the segments' positions and shapes. First, the wave fronts of the segments must be measured with a WFS, which we wish to avoid. Stellar interferometry had shown us that the fringes between every two segments in the aperture must be unique in order to identify them. These two segments contribute to two symmetric and distinct side lobes ("base line") in the Optical Transfer Function (OTF) or its modulus (MTF). The OTF is the autocorrelation of the pupil map, and is obtained by Fourier transforming the Point Spread Function (PSF), where the fringes appear. However, each position in the MTF corresponds to all equal base lines. In other words, to identify one such pair, these base lines must be non-redundant [20-22]. The inclusion of redundant base lines does not contribute to the resolution, only to the signal-to-noise ratio. Current research on further aspects of sparse aperture imaging deals mainly with equal and symmetrical segments, such as circles and hexagons [22-24].

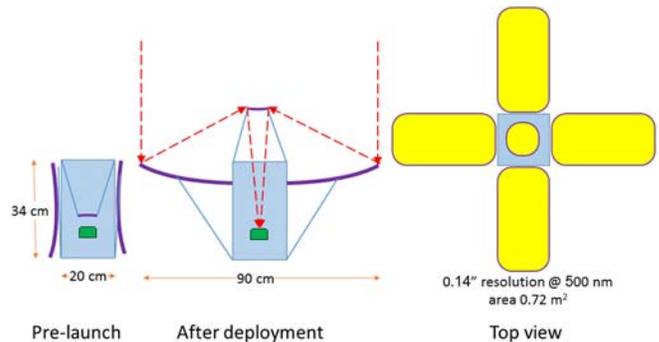

**Fig. 1.** Concept of a basic deployable telescope.

To break segment symmetry, we devised an optimized search programme, to place the segments around a central hole, in a Newtonian set-up (Figure 2). These sectors were cut out of a parabolic mirror, and for mechanical simplicity we used only four. Notice that now the base line 1-2 is distinct from the 3-4 base line, as opposed to Figure 1. The OTF of a point source, as seen here, is the deconvolution kernel for images obtained with the corrected segmented sparse telescope.

Here we see the other main problem with the design in Figures 1 and 2: the straight edges of the segments scatter a great deal of light, evident as radial spokes in the PSF (Figure 2c), and inhibit the detection of faint objects near a bright one. Thus to minimize diffraction, we tried different shapes of rounded segments, from circular to elliptical. Ellipses are also advantageous from the manufacturing point of view, for mechanical and thermal stability of the telescope, as well as for diffraction suppression from straight

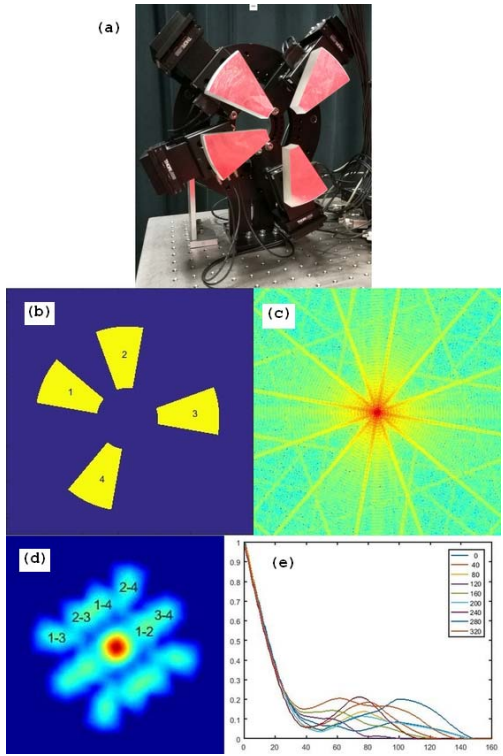

**Fig. 2**. (a) A laboratory model of a 20 cm sparse telescope (reflecting off a red sheet); (b) The sectors are set at non-redundant angles (rotated with respect to (a); (c), The calculated PSF, with strong scattering off the straight sector edges (log scale); (d) The central part of the MTF, with limited gaps between the side lobes.; (e) Radial cuts in the MTF.

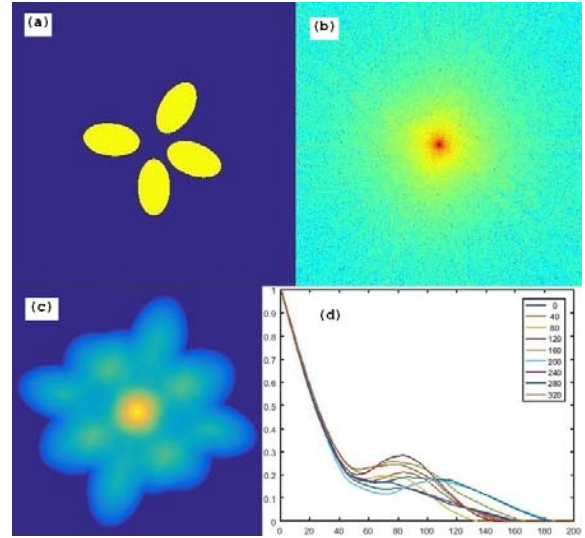

**Fig. 3**. (a) Elliptical segments, in a typical realization, to be compared with Fig. 2; (b) The PSF (log scale); (c) The central part of the MTF. No gaps exist between the side lobes.; (d) Radial cuts in the MTF.

and edge boundaries. In our case, a shifted and rotated ellipse $\varepsilon_\alpha$ is defined by:

$$(x'\cos\alpha + y\sin\alpha)^2/a^2 + (x'\sin\alpha - y\cos\alpha)^2/b^2 \leq 1; x' \equiv x - ka \quad (1)$$

where $a, b$ are the ellipse semi-axes, and $ka = 1.5a$ is the off-center radial shift of the segment before rotation about the center. The aperture function is $P = \Sigma_{i=1,4}\varepsilon_i$ where the ellipses $\varepsilon_i$ have different rotation angles $\alpha$ and are pairwise disjoint. The pupil auto-correlation function is $Q = P \otimes P$, its area is $A$ (where $|Q| > 0$) and its standard deviation is $S = STD(Q)$. We ran a fully automated optimisation with the axes ratio $a/b$ of the segments as one free parameter, and their angular location around the telescope aperture as three more parameters. The first target criterion was the widest MTF (maximal area $A$), namely with the broadest $uv$ (Fourier frequencies) coverage. However, narrow gaps sometimes crept among the side lobes, and these solutions were excluded. We tried another optimisation, namely maximizing the MTF smoothness $S$, and rejecting solutions with gaps in their MTF, so as to avoid zero division in Wiener filtering of the images [12, 22]. The results for the two criteria were very similar.

We ran many trials with multiple solutions, which we sorted to correct for the rotation and reflection symmetries (Figure 3). All configurations had figures of merit, either MTF area or smoothness, which varied only by a few percent. However, the ellipticity always turned out to be either maximal or minimal, according to the optimisation criterion. Since packaging constraints will settle the final ellipticity, we set first the axes ratio, then optimized only the angles among the segments. For the axes ratio $a/b$ of 1.5 to 2.2, the azimuthal angles between segments were 66.8±3.4, 83.8±5.2, 113.0±4.5 and 96.4±6.3 degrees. At any rate, the rounder edges produced a more regular PSF than straight ones (compare Figures 2e and 3d). Numerically, the trends were weak and opposite: for the same $b/a$ range of 0.45 to 0.67, the autocorrelation area rose linearly from 0.61 to 0.69, relative to the full aperture area. The smoothness had the same weak behaviour, rising from 0.112 to

0.122. This means that broader autocorrelation was also lumpier, since longer ellipses produced more serrated MTF edges.

Next we needed to resolve two alignment issues, correcting for tip-tilt errors of the segments, and for their piston phasing (equalisation of their paths). From our own experience in the past, we knew that correcting for tip-tilt is the easier problem. We have used simulated annealing [5-7] to optimize the PSF and maximise the image sharpness. We have found that for segmented telescopes, correction for tip-tilt was fast, but phasing was very slow, essentially an asymptotic convergence process.

We assumed that the focus errors on the separate segments were small. When they are large, we will need to take few images in different focus positions to find the correct distance for each segment, which happens when each PSF is the sharpest [13], then move them to the right position. (We assume that the camera field-of-view is large enough to capture all four PSFs). To identify each segment, one can tilt sequentially the sections and identify them [13, 25]. We opted to use the broken rotational symmetries of the ellipses [14] which does the same without motion (Figure 4): As the segments have known orientations, they also have distinct PSFs, which can be measured directly by the camera. In our simulation we calculated four such PSFs for the segments, which simply amount to rotated elliptical Airy discs. Then we cross-correlated them with the spots in the image plane of the telescope (Figure 4b). The shifts in the correlations maxima indicated the tip-tilt errors of the segments, down to a fraction of a wave length. In some cases, where the PSFs overlapped, the correlations tended to fail. In these cases, we backed up from the focal plane to a defocus position (Figure 4c). Here we could achieve the separate Fresnel diffraction patterns of the segments, also unique and elliptical. Next we applied the same correlation location algorithm, and removed the ambiguity between the elements. Notice that no two segment PSFs can overlap both in focus and out-of-focus. This accomplished the angular alignment of the segments, a stage also known as "stacking".

After the tilt on the segments was minimized, the last and longest step was to phase ("piston") them to be at the same distance from focus, within a few wave lengths. We assumed that the initial errors in the tip, tilt, and piston, were up to ±1 mm, and the working wave length was in the visible, ~0.5 μm. Now we employed a method akin to optical coherence tomography (OCT): we scanned one segment and observed the combined PSF of all segments, to see when it varied in intensity [1, 11, 15]. Such a variation meant that fringes ran through the image in focus. We also assumed that the observed object is a bright star of white spectrum. This limited very much the coherence length of the beam, so fringes were only visible for a few wave lengths before and after they had equal paths among the segments. Narrow flip-in spectral filters could increase this range, but we sought to eliminate additional mechanical components, so as to maintain system reliability in space. For earth-observing and for astronomical telescopes we assumed using an unresolved star for the initial alignment stages. We have previously aligned successfully a 7-DOF system on an extended ground scene in the lab, but optimization during short fly-over time was difficult [11].

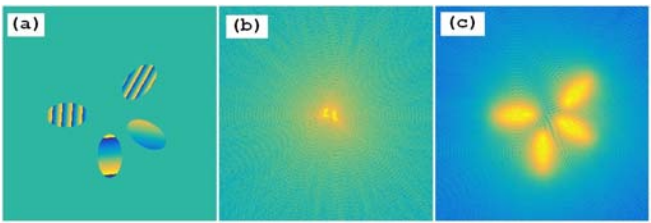

**Fig. 4.** (a) Tilted elliptical segments, showing wrapped phase (different realization from Figure 3); (b), The resultant PSF in the center of the camera (log scale); and (c) Slightly defocused image (log scale).

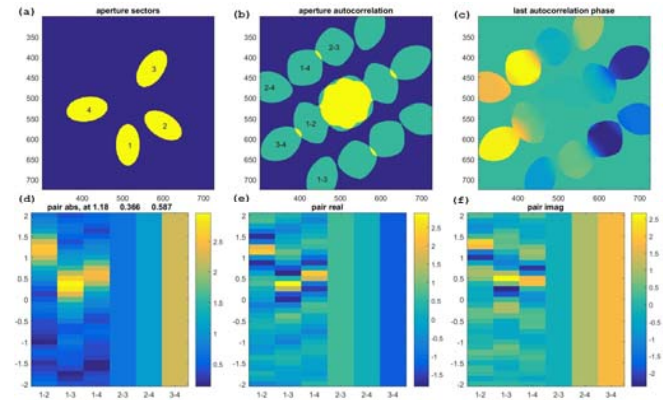

**Fig. 5.** Short-scan example, segment 1 moving: (a) Numbered segments; (b) MTF, showing side lobes corresponding to segment pairs (excluding the symmetric conjugate ones), threshold ~5%; (c); The last phase of the OTF at position 2 mm; (d-f) Amplitude, real and imaginary parts of the OTF, sampled at the centers of the side lobes in (b), before noise addition. The side lobes were well-separated, so fringes were not visible in pairs that did not involve segment 1 (flat vertical stripes).

Since the segments were placed at non-redundant positions, we identified the two segments that contributed to the fringes. An example is shown in Figure 5. As we moved segment 1, we could see fringes crossing the side lobes corresponding to the other three segments, pairs 1-2, 1-3, 1-4. These fringe crossings happened each at a different scan position, according to the optical path differences with the other three segments. In some simulations we moved all four segments sequentially, and measured their mutual interferences. However, we only looked for the differences between the four segments, so the multiple scans increased the search time by a factor of four, with the only benefit of better noise figure. A least-squares solution gave the relative piston values [16]. At this simulation stage, no use was made of the constraint of zero phase closure [17], which might have improved performance even further.

Using a 1024 x 1024 array, we chose step sizes from six to eight to the micrometer, and added some position noise to simulate the lack of feedback on the actual position. The wave lengths of the simulations were spread between 0.37 μm and 0.67 μm; 12-15 wave lengths were chosen, in uneven spacings, to minimize mutual beating and aliasing in the simulations. After the values of the complex OTF were sampled at the center pixels of the side lobes, Gaussian noise was added to them, both real and imaginary. This is because the Poisson and detector noise spread evenly in the Fourier domain. As long as the signal-to-noise ratio exceeded 8, the position accuracy of the sectors was down to 50 nm. Even higher accuracy could be obtained by also using the real part of the OTF (Figure 5e). In addition, more area of all twelve side lobes could be used to better determine the fringes (Figure 5c), except where they overlap with the next side lobe (Figure 5b). Thus the non-redundant masking demands can be relaxed in some cases: the segments sizes can be increased, enlarging also the overlap area, as long as we have at least

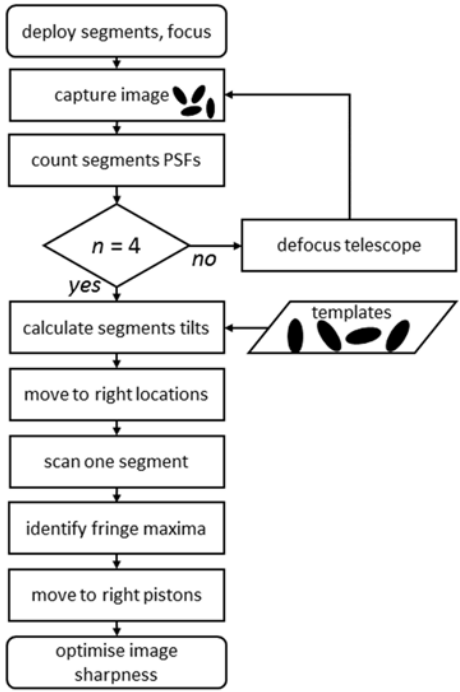

**Fig. 6**. Flow-chart of the process of correcting for tip-tilt and then piston of the telescope segments.

one point in each side lobe in the OTF which is unique to its pair of segments only [15].

We reached the correct tip, tilt and piston values to within ~1 µm (Figure 5d), since we only knew the average step size, and have assumed no feedback on the actuator positions. Henceforth, simple optimization was sufficient to reach the best solution. Other methods for the final fine correction could also be used [13, 18, 19].

The whole alignment process is summarized in Figure 6. It is to be compared with the optimization alternative [7]. In that optimization, we tried different methods, such as gradient descent, simulated annealing, and stochastic parallel gradient descent, a combination of the two. While simulated annealing is proven to have a solution, even if very slow to arrive at, the other methods tend to sink in local minima and are thus not reliable. In our lab setup, we have 17 DOFs (tip, tilt, fine and coarse piston for each segment, and global focus), where each of these has about 1,000 gradations. This makes the full search volume for general optimization methods enormous: $10^{51}$ options. But when starting near the best solution, and dropping coarse pistons and focus DOFs, we get down to ~$10^{19}$ options. With the chances of multiple minima considerably reduced, gradient search should suffice.

The week time limit for the full alignment is quite weak. This is because a new space entity anyway requires days to initially set itself up. Still, it is beneficial to limit the search volume, which is what we propose here. Segment alignment can take minutes, and scanning one segment against the others, while searching for mutual fringes, can take tens of minutes: the range is a few millimeters, and the step size – a fraction of a micron. But by the time these two main steps are over, the segments are within a micron from their optimal position, search should be faster. Notice that these search procedures are not limiting: if they fail (for example by a non-functioning actuator), simulated annealing can still optimize the image sharpness, from scratch, with the constraint on the bad component included indirectly.

Some of the assumptions we have made are going to be tested in the lab, such as unmetered actuator motion and wide band light. When using simulated annealing optimization, calibrated step sizes or fully decoupled motions are not important, as opposed to gradient search methods. (The only exception is the focus motor, which accurate position is required for later operation of the telescope, e. g. to correct for thermal distortions of the structure). Erroneous assumptions about the locations of the motors or actuators are more important in the second stage, of phasing the elements, since long motions, with attendant larger errors, are unavoidable. However, the search scheme (Figure 6) can be repeated if the image quality is insufficient before starting the final optimization stage.

To summarise, we have made use of both the non-redundant shape of the aperture segments, and the non-redundant placement of the segments, to tell them apart, and to identify the segment pairs. The search volume for the best segment alignment was significantly reduced, and with it the total alignment process time. All of these were achieved with a minimally complex space telescope, and with the least assumptions on mechanical or spectral precision.

# References

1. J. M. Beckers, E. K. Hege, and P. A. Strittmatter, Optical interferometry with the MMT, *Proc. SPIE* **444**, 85-92 (1983).
2. G. A. Chanan, M. Troy, and C. M. Ohara, Phasing the primary mirror segments of the Keck telescopes: a comparison of different techniques, *Proc. SPIE* **4003,** 188-202(2000).
3. R. A. Muller and A. Buffington, Real-time correction of atmospherically degraded telescope images through image sharpening, *J. Opt. Soc. Am.* **64**, 1200-1210 (1974).
4. J. P. Hamaker, J. D. O'Sullivan, and J. E. Noordam, Image sharpness, Fourier optics, and redundant-spacing interferometry, *J. Opt. Soc. Am.* **67**, 1122-1123 (1977).
5. E. N. Ribak, J. Adler, and S. G. Lipson, Telescope phasing and ground states of solid-on-solid models. *J. Phys. A* **23**, L809-L814 (1990).
6. S. Zommer, E. N. Ribak, S. G. Lipson, and J. Adler: Simulated annealing in ocular adaptive optics, *Opt. Lett.* **31**, 939-941 (2006).
7. I. Paykin, L. Yacobi, J. Adler, and E. N. Ribak, Phasing a segmented telescope, *Phys. Rev. E* **91**, 023302 (2015).
8. X. Ma, Z. Xie, H. Ma, Y. Xu, G. Ren and Y. Liu, Piston sensing of sparse aperture systems with a single broadband image via deep learning. *Opt. Express* **27**, 16058-16070 (2019).
9. M.R. Bolcar and J.R. Fienup, Sub-aperture piston phase diversity for segmented and multi-aperture systems, *Appl. Opt.* **48,** A5-A12 (2009).
10. I. Surdej, N. Yaitskova, and F. Gonte, On-sky performance of the Zernike phase contrast sensor for the phasing of segmented telescopes. *Appl. Opt.* **49**, 4052-4062 (2010).
11. D. Dolkens, G. Van Marrewijk and H. Kuiper, Active correction system of a deployable telescope for Earth observation. ICSO 2018, *Proc. SPIE* **11180**, 111800A (2019).
12. M. J. Golay, Point arrays having compact, nonredundant autocorrelations. *J. Opt. Soc. Am.* **64,** 272-273 (1971).
13. A. B. Meinel and M. P. Meinel, Optical Phased Array Configuration for an Extremely Large Telescope, *Applied Optics* **43**, 601–607, (2004).
14. J. B. Breckinridge, Sparse-Aperture Telescopes. Chapter 13 in *Basic Optics for the Astronomical Sciences*. SPIE Digital Library (2012).
15. F. Cassaing and L. M. Mugnier, Optimal sparse apertures for phased-array imaging, *Opt. Lett*. **43**, 4655-4658 (2018).
16. P. Tuthill, Masking interferometry at 150: old enough to mellow on redundancy? *Proc. SPIE* **10701**, 107010S.
17. J. R. Fienup, D. K. Griffith, L. Harrington, A. M. Kowalczyk, J. J. Miller and J. A. Mooney, Comparison of reconstruction algorithms for images from sparse-aperture systems. *Proc. SPIE* **4792**, 1-8 (2002)
18. D. S. Acton, J. S. Knight, A. Contos, S. Grimaldi, J. Terry, P. Lightsey, A. Barto, B. League, B. Dean, J. S. Smith, C. Bowers, D. Aronstein, L. Feinberg, W. Hayden, T. Comeau, R. Soummer, E. Elliott, M. Perrin, and C. W. Starr Jr., Wavefront Sensing and Controls for the James Webb Space Telescope, *Proc. SPIE* **8442**, 84422H (2012).
19. E. N. Ribak and S. Gladysz, Fainter and closer: finding planets by symmetry breaking, *Opt. Express* **16**, 15553-62 (2008).
20. T. J. Schulz and R. G. Paxman, Piston alignment for a segmented-aperture imaging system by using piston-sweep phasing. *Opt. Lett*. **42**, 2922-2925 (2017).
21. A. Z. Greenbaum, L. Pueyo, A. Sivaramakrishnan, and S. Lacour, An image-plane algorithm for JWST's non-redundant aperture mask data. *Astroph. J.* **798**, 68 (2014).
22. W. Zhao and Q. Zeng, Simultaneous multi-piston measurement method in segmented telescopes. *Opt. express* **25,** 24540-24552 (2017).
23. M. J. Ireland, D. Defrère, F. Martinache, J. D. Monnier, B. Norris, P. Tuthill, and J. Woillez, Image-plane fringe tracker for adaptive-optics assisted long baseline interferometry, *Proc. SPIE* **10701**, 1070111 (2018).
24. S. Vievard, F. Cassaing, and L. M. Mugnier, Large amplitude tip/tilt estimation by geometric diversity for multiple-aperture telescopes. *J. Opt. Soc. Am. A* **34**, 1272-1284 (2017).
25. L. H. Huang, Coherent beam combination using a general model-based method. *Chinese Physics Letters* **319**, 094205 (2014).